\documentclass[twocolumn,prb,aps,showpacs]{revtex4-1}
\usepackage{graphicx}
\usepackage{color}
\usepackage{dcolumn}
\usepackage{amsmath}
\usepackage{units}

\newcommand{\TC}{T$_c$ }
\newcommand{\EF}{E$_F$ }
\newcommand{\FB}{FeSe$^{BL}$ }
\newcommand{\FS}{FeSe$^{SL}$ }

\begin{document}

\title{20 K superconductivity in heavily electron doped surface layer of FeSe bulk crystal}

\author{J. J. Seo$^1$, B. Y. Kim$^1$, B. S. Kim$^{2,3}$, J. K. Jeong$^1$, J. M. Ok$^4$, J. S. Kim$^4$, J. D. Denlinger$^5$, C. Kim$^{2,3,*}$, Y. K. Kim$^{2,3,5,*}$ }

\address{$^1$Institute of Physics and Applied Physics, Yonsei University, Seoul 120-749, Korea}
\address{$^2$Center for Correlated Electron Systems, Institute for Basic Science, Seoul 151-742, Korea}
\address{$^3$Department of Physics and Astronomy, Seoul National University (SNU), Seoul 151-742, Korea}
\address{$^4$Department of Physics, Pohang University of Science and Technology, Pohang 790-784, Korea}
\address{$^5$Advanced Light Source, Lawrence Berkeley National Laboratory, Berkeley, CA 94720, USA}

\maketitle

\noindent

\vspace{10 pt}
\vspace{10 pt}

{\bf A superconducting transition temperature (T$_C$) as high as 100 K was recently discovered in 1 monolayer (1ML) FeSe grown on SrTiO$_3$ (STO)\cite{Wang, Zhang, Deng, Liu, LiuX, JF}. The discovery immediately ignited efforts to identify the mechanism for the dramatically enhanced \TC from its bulk value of 7 K. Currently, there are two main views on the origin of the enhanced T$_C$; in the first view, the enhancement comes from an interfacial effect while in the other it is from excess electrons with strong correlation strength. The issue is controversial and there are evidences that support each view. Finding the origin of the T$_C$ enhancement could be the key to achieving even higher T$_C$ and to identifying the microscopic mechanism for the superconductivity in iron-based materials. Here, we report the observation of 20 K superconductivity in the electron doped surface layer of FeSe. The electronic state of the surface layer possesses all the key spectroscopic aspects of the 1ML FeSe on STO. Without any interface effect, the surface layer state is found to have a moderate \TC of 20 K with a smaller gap opening of 4 meV. Our results clearly show that excess electrons with strong correlation strength alone cannot induce the maximum \TC, which in turn strongly suggests need for an interfacial effect to reach the enhanced \TC found in 1ML FeSe/STO.}

\begin{figure}[t!]
\hspace*{-0.2cm}\vspace*{-0.1cm}\centerline{\includegraphics[width=1\columnwidth,angle=0]{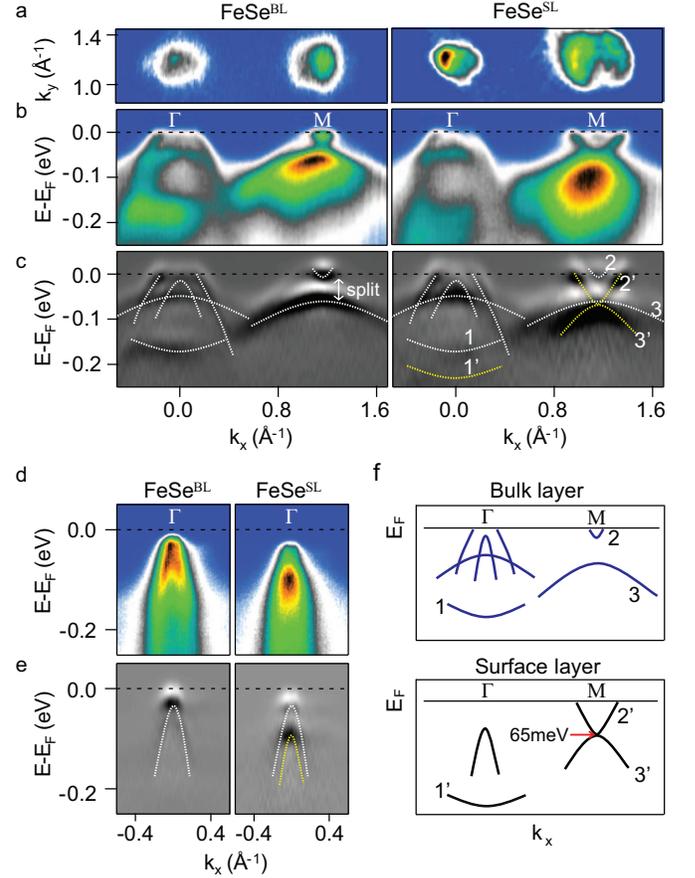}}
\caption{{\bf Electronic structures of pristine and surface electron doped FeSe.} ({\bf a}) Fermi surface mapping of pristine (FeSe$^{BL}$) and surface doped (FeSe$^{SL}$) FeSe, measured at 30 K. ({\bf b}) Band dispersions along the $\Gamma$-M high symmetry line of \FB (left) and  \FS (right), and ({\bf c}) Second derivatives of ({\bf b}). White and yellow dashed lines indicate the band dispersions of \FB and \FS, respectively. ({\bf d}) Band dispersion around the $\Gamma$-point in a different geometry for \FB (left) and \FS (right). ({\bf e}) Second derivatives ({\bf d}). ({\bf f}) Schematic of the band dispersions of \FB (upper) and \FS (lower).}
\end{figure}

A strikingly enhanced \TC, far above the previous record of \TC in bulk iron based superconductors, was discovered in a relatively simple system of 1ML FeSe on STO. The observation quickly initiated extensive and intensive studies to unveil the key mechanism for the enhancement. The mechanism, if found, should be important in its own right but may also provide key information on the superconducting mechanism in iron based superconductors.

\begin{figure}
\hspace*{-0.2cm}\vspace*{-0.1cm}\centerline{\includegraphics[width=1\columnwidth,angle=0]{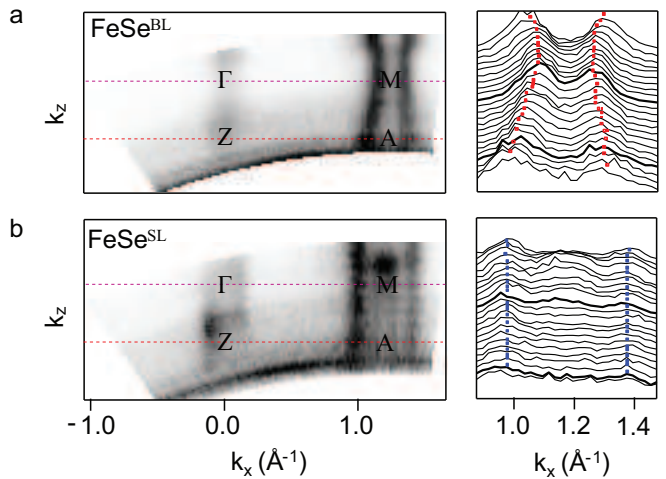}}
\caption{\textbf{k$_z$ dependence of \FB and \FS.} ({\bf a}) Constant energy map in the k$_x$-k$_z$ plane from \FB normal state at 120K (left) and stacked MDCs near the M-point (right). ({\bf b}) k$_z$ dependence of \FS at 30 K.}
\vspace*{-0.5cm}
\end{figure}

\begin{figure}
\hspace*{-0.2cm}\vspace*{-0.1cm}\centerline{\includegraphics[width=1\columnwidth,angle=0]{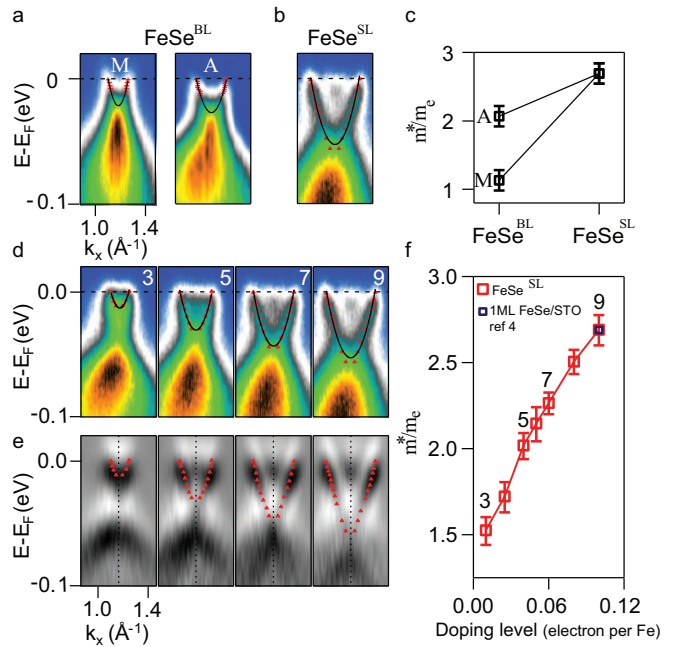}}
\caption{\textbf{Evolution of the effective mass in \FS.} ({\bf a}) Electron band near the M-point at k$_z$=0 ($\pi$) before surface electron doping. Overlaid on each plot is a parabolic fit used in extracting the effective mass. ({\bf b}) The same cut after surface electron doping. ({\bf c}) Effective masses of the \FB and \FS at different $k_z$. ({\bf d}) Evolution of the electron band with surface electron doping. Parabolic fits of the peak positions (red triangles) are overlaid. ({\bf e}) Second derivatives of ({\bf d}). Red triangles are peak positions. ({\bf f}) Evolution of the effective mass from \FB to \FS as a function of surface doping level.}
\vspace*{-0.5cm}
\end{figure}

\begin{figure*}
\hspace*{-0.2cm}\vspace*{-0.1cm}\centerline{\includegraphics[width=1.8\columnwidth,angle=0]{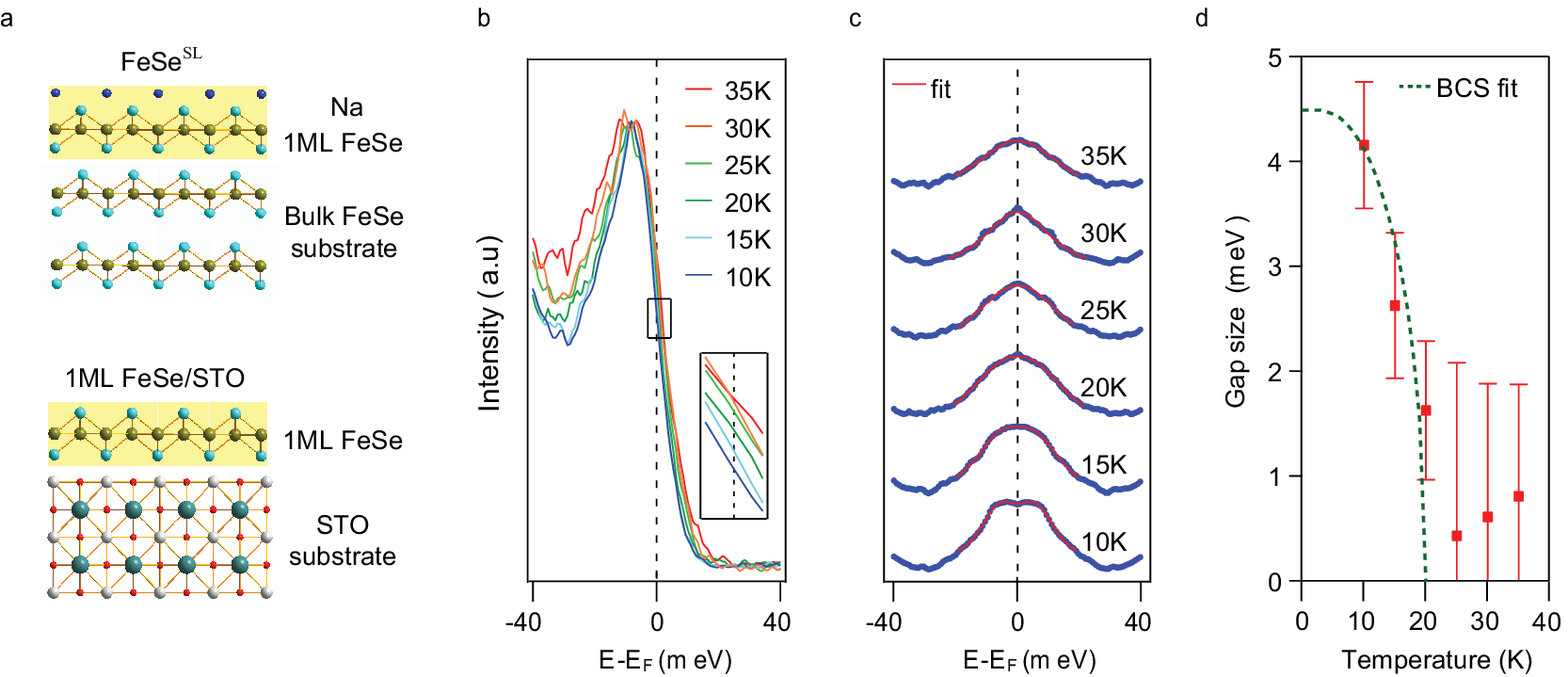}}
\caption{\textbf{Superconducting gap of the \FS state.} ({\bf a}) Schematic views of the induced surface layer state \FS (upper) and 1ML FeSe/STO (lower). ({\bf b}) Temperature dependent EDCs taken at the Fermi momentum of the electron pocket at the M-point. ({\bf c}) Symmetrized EDCs of ({\bf b}). Each spectrum is fitted with a Dynes function and the result is overlaid as a solid red curve. ({\bf d}) Temperature dependent superconducting gap size extracted from the fit. The dashed green line is the BCS gap function.}
\vspace*{-0.5cm}
\end{figure*}

Two views are mainly considered on the issue at present. In the first view, the origin of the enhancement comes from the FeSe layer. Angle resolved photoemission spectroscopy (ARPES) studies have shown that 1ML FeSe on STO is heavily electron doped with electrons provided by the substrate and, as a result, has only electron pockets in the Fermi surface\cite{He, Tan}. The observed electron bands are also found to have insulator-superconductor crossover with an enhanced electron correlation strength which is possibly due to confinement of electrons in two-dimensional state or strain from the substrate\cite{Wenhao, Junfeng}. It was then proposed that 1ML FeSe/STO shares the same superconducting mechanism with ordinary iron-based superconductors which are considered to be strongly correlated electron systems as cuprates\cite{SiQ, Dai, Yin}.

In the other view, the origin comes from outside of the FeSe layer. That is, a strong interfacial effect is an essential ingredient of the large \TC enhancement\cite{Yun, Lee, Cui, Miyata, Tan}. This view is based on the fact that the enhanced \TC is observed only near the interface\cite{Tan, Miyata}. As for what exactly the interface effect is, two possibilities have been raised so far. The first one is an additional pairing channel provided by the STO phonons\cite{Lee}. The observation of a replica band, believed to be a good of strong coupling between an electron in the FeSe layer and an optical phonon mode of the underlying STO, suggests a significant role of such additional pairing channel\cite{Lee}. The other possibility comes from the stabilization of an ordered state by the interface which should provide strong spin fluctuation when it is broken by electron doping. This view is based on the observation in an earlier experiment that the phase transition temperature increases with less number of layers\cite{Tan}.

So far, there is no experimental evidence that can clearly resolve the issue. A simple way to resolve the issue would be to fabricate a free standing 1ML FeSe with excess electrons and see at which temperature system becomes superconducting. It can clearly tell us which factor is dominant for the enhancement. However, it is practically impossible to achieve. Instead, our strategy is to closely mimic the situation by inducing a monolayer-like FeSe state on a FeSe bulk crystal. It was induced via surface electron doping which can be done by alkali metal evaporation\cite{Hossain, YKKim}. In the electronic structure point of view, the induced state possesses all the key characteristic aspects of 1ML FeSe/STO: i) heavy electron doping, ii) reduced dimensionality, and iii) enhanced electron correlation strength. Thus the induced state is almost identical to 1ML FeSe/STO. The only difference is any source of interfacial effect is not expected. Therefore, the resulting \TC enhancement in the induced state can tell us what is the main ingredient for the giant \TC enhancement in 1ML FeSe/STO. We now demonstrate how does the induced state on bulk FeSe follow those three characteristics of 1ML FeSe and show how \TC enhanced.

\noindent
\newline
{\bf Results}
\newline
{\bf Electronic structures and surface electron doping.} We first show that the doping level achieved via surface electron doping can reach that of the 1ML FeSe/STO. Figure 1b shows the band dispersions along the $\Gamma$-M high symmetry line of pristine (FeSe$^{BL}$) and surface doped (FeSe$^{SL}$) samples measured at 30 K. The electron band in \FS has a downward shift with a larger Fermi surface. The observed shift, judging from the electron band bottom location of 65 meV, is similar to the value for 1ML FeSe/STO\cite{Liu}, was kept to make same doping level. Based on the calculated Luttinger volume, there are 0.1 excessive electrons per Fe, similar to the doping level of 1ML FeSe/STO (0.1$\sim$0.12) that makes maximum \TC (55$\sim$65 K)\cite{Liu, He}. As for the hole band, it first looks as if there is not much change in the dispersion upon surface doping. However, a close inspection of the data taken with various geometries shows a downward shift of the hole band (see Figs. 1d and 1e). A tiny and faint electron band at the M-point still remains with the size very close to that of the pristine sample. The observation of both surface and bulk states can be understood to be from different length scales of the charge doping and probing depth. That is, the probing depth of ARPES is larger than the charge doping depth, and as a consequence signals from both the doped surface and underlying bulk states are seen.

A notable aspect of the band dispersion in \FS is that it does not have the split bands near the M-point which are believed to be a manifestation of the ferro-orbital ordering\cite{Shimojima, Nakayama, Baek}. This suggests that ferro-orbital ordering is suppressed through the surface electron doping. With both the hole and electron bands simply shifted to the higher binding energy side and the ferro-orbital ordering suppressed, the overall band dispersion of \FS fully replicates that of the 1ML FeSe/STO. The full band assignments are made with the second derivative data in Fig. 1c and 1e, and the results are summarized in Fig. 1f for \FB (upper) and \FS (lower), respectively.

\noindent
\newline
{\bf Low dimensionality of the induced state.} We next show that the doping induced state on the surface is almost two dimensional. If the state is two dimensional, there is no out-of-plane momentum (\emph{k}$_\text{z}$) dependence (or photon energy dependence in the experiment) in the band structure. Figure 2a shows Fermi surface maps of FeSe$^{BL}$ and FeSe$^{SL}$ in the \emph{k}$_\text{z}$-\emph{k}$_\text{x}$ plane. The data for FeSe$^{BL}$ was taken at 120 K to avoid complications from the ordered phase while FeSe$^{SL}$ data was taken at a lower temperature of 30 K. The figure also has stacked momentum distribution curves (MDCs) near the M-point. \FB has a weak but clear three-dimensional electronic structure modulation in both hole and electron Fermi surfaces. On the other hand, the \FS case given in Fig. 2b shows no modulation along the \emph{k}$_\text{z}$ direction. This implies that the state is confined within a two-dimension layer or, at least, it has negligibly weak inter-layer interaction. We conclude the former is the case, in detail, only the first layer of FeSe is doped.

Another characteristic feature of the two-dimensional confinement is the strengthening of the electron correlation\cite{Junfeng, LEEPA, Stewart, Brinkman, Reyren, Richter}. A way to examine the electron correlation strength is to check the effective mass. The effective mass can be obtained from a parabolic fit of the experimental band dispersion in Fig. 3a and 3b. As shown in Fig. 3c, the effective mass of \FS state is m*/m$_e$=2.7, larger than that of \FB (m*/m$_e$=1.1 and 2.1 at k$_z$=0 and $\pi$, respectively). It clearly indicates a stronger electron correlation in \FS.  Furthermore, the value is consistent with 1ML FeSe/STO (m*/m$_e$=2.7)\cite{Liu}. We also note that the evolution of the effective mass upon surface electron doping in Figs. 3e, 3d and 3f shows a gradual increase without any abrupt change. It strongly suggests that no phase transition is involved in the observed change in the effective mass.
%Figure 2c shows electron band dispersions of \FS around zone corner at different $k_z$=$0$ (left), $\pi$ (right). That of \FS is given in Fig. 2d

\noindent
\newline
{\bf Confinement $\&$ enhanced correlation.}  First evidence for a single layer doping is that no multilayer band splitting is observed.  If electron permeates into several layers with a potential gradient, then the band should be broad since we will then measure the sum of bands with different doping level. However, the fact that we only observed two types of band - surface induced state \FS and unaffected state \FB with the sharp and clear features as shown in the spectra, supports our view. Another evident feature of the two-dimensional confinement is the strengthening of the electron correlation\cite{Junfeng, LEEPA, Stewart, Brinkman, Reyren, Richter}. A way to examine the electron correlation strength is to check the effective mass. As shown in Fig. 3c, the effective mass of \FS state is \emph{m}$^*$/\emph{m}$_\text{e}$=2.7, larger than that of \FB (\emph{m}$^*$/\emph{m}$_\text{e}$=1.1 and 2.1 at \emph{k}$_\text{z}$=0 and $\pi$, respectively). It clearly indicates a stronger electron correlation in \FS. The value is consistent with that observed in 1ML FeSe/STO (m*/m$_e$=2.7)\cite{Liu}.
We note that the evolution of the effective mass upon surface electron doping in Figs. 3e, 3d and 3f shows a gradual increase without any abrupt change. It strongly suggests that no phase transition is involved in the observed change in the effective mass. Also as we cannot expect the any strain applied in the system, the effective mass enhancement is sorely due to dimensionality reducing, i.e. confinement. With all these evidences, we conclude that the doped electrons reside almost within the first layer of FeSe possibly due to the weak van der Waals coupling between FeSe layers.

\noindent
\newline
{\bf Superconducting gap.} We have demonstrated that  \FS mimics three spectroscopic traits of 1ML FeSe/STO. The next and important step is to check how the \TC changes in the induced state. Figure 4 shows the result of superconducting gap measurements. Leading edge shift upon cooling is captured in the raw EDCs from the Fermi momentum of the electron band (see Fig. 4b). Symmetrized EDCs given in Fig. 4c, in which the the influence from the Fermi-Dirac distribution function is eliminated, show a gap feature at the lowest temperature with a size of 4 meV ($\pm$ 0.6), obtained by fitting the data with a Dynes function\cite{Dynes}. Temperature dependence of the gap size roughly traces BCS type order parameter dependence with a \TC of 20 K, which results in 2$\Delta$/k$_B$T$_C$ of 5.22. This is somewhat smaller than that of 6-7 for 1ML FeSe/STO\cite{He}.

\noindent
\newline
{\bf Discussion}
\newline
So far, we have shown that \FS possesses all characteristic traits of 1ML FeSe/STO: heavy electron doping, two-dimensional state within a layer, and enhanced electron correlation. Fact that additional electrons from evaporated Na undergo only to the first top layer make it possible. Reduced dimensionality was reflected in the absence of k$_z$ modulation and in the enhanced effective mass that indicates the enhanced electron correlation by the confinement of electrons in two-dimension. Doping level was also kept to similar with 1 ML FeSe/STO. Therefore \FS is identical to that in 1 ML FeSe/STO, but FeSe layer side only. A clear distinction is, \FS is free from any possible interface effects unlike 1 ML FeSe on STO case. This distinction, together with the observed 20 K superconductivity in \FS that far lower than what is observed in 1ML FeSe/STO, leads our conclusion that heavy electron doping with enhanced electron correlation can increase the \TC only to a limited value.

It was very recently reported that surface doped 30ML FeSe film with K-dosing has a \TC of about 40 K\cite{Wen}. We believe that only the top layer of their thin film was also doped, considering our findings. In addition, Li$_{0.84}$Fe$_{0.16}$OHFe$_{0.98}$Se was reported to have two-dimensional electronic structure with a \TC of 41 K because of the enlarged spacing between FeSe layers from (LiFe)OH intercalation\cite{Zhao}. Even though it is still to be understood why the \TC of our surface doped bulk FeSe is lower than other systems, all the results including ours point to our earlier statement that additional electron doping with strong correlation can enhance the superconductivity but only to a limited value of around 40 K. It therefore clearly suggests a need of interface effects for realizing the highest \TC observed only in 1ML FeSe/STO. The exact role of interface effect is not clear yet. Further efforts are desired to unveil the role of interface effect.

\noindent
\newline
\textbf{Methods}
\newline
\textbf{ARPES measurement} ARPES measurements were performed at the Beamline (BL)10.0.1 and 4.0.3 of the Advanced Light Source. Surface electron doping was done by Na evaporation on the sample surface using commercial SAES alkali metal dispensers. Spectra were taken with Scienta R4000 (BL 10.0.1) and R8000 (BL 4.0.3) electron analyzers with overall energy resolutions of 10 meV (BL 10.0.1) and 13 meV (BL 4.0.3), respectively. Photon energy dependent measurements were performed with photon energies from 50 to 90 eV. The samples were cleaved and doped at 30 K in an ultra-high vacuum better than 4X10$^{-11}$ Torr. All measurements were performed within one hour per sample because of the short surface life time after Na evaporation.
\newline
\textbf{Data analysis} Luttinger volume, electron pocket area was calculated by assuming two elliptical Fermi surfaces perpendicular to each other with \emph{k}$_\text{F}$: a = 0.46 $\AA^{-1}$ , b = 0.38 $\AA^{-1}$ extracted from high symmetry line band dispersion. The spectra symmetrization in Fig. 4c was done by reverting the spectra with respect to \EF after being divided by the corresponding Fermi Dirac distribution function. BCS fit in Fig. 4d was calculated using temperature dependent BCS gap function under weak-coupling limit \emph{E}$_\text{g}$(T) = \emph{E}$_\text{g}$(0) tanh ( $\frac{\pi}{2}$$\sqrt{\frac{Tc}{T}-1}$ )\cite{Gross, Inosov}. Dynes fit for superconducting energy gap was referenced at ref 29.
\noindent
\newline

\noindent
\newline
{\bf Acknowledgements}
\newline
This work is supported by IBS-R009-G2 through the IBS Centre for Correlated Electron Systems. The Advanced Light Source is supported by the Office of Basic Energy Sciences of the U.S. DOE under Contract No. DE-AC02-05CH11231. The work at POSTECH was supported by the NRF through SRC (Grant No. 2011-0030785) and Max Plank POSTECH/KOREA Research lnitiative (Grant No. 2011-0031558) programs, and also by IBS (No. IBSR014-D1-2014-a02) through Center for Artificial Low Dimensional Electronic Systems.

\noindent
\newline
{\bf Author Contributions}
\newline
J.M.O. and J.S.K. grew the crystals; J.J.S., B.Y.K., B.S.K. and J.K.J. performed ARPES measurements; J.J.S. analysed the ARPES data; J.J.S., C.K. and Y.K.K. wrote the paper; Y.K.K. and C.K. are responsible for project direction and planning.

\noindent
\newline
{\bf Additional information}
\newline
Reprints and permissions information are available online at www.nature.com/reprints. Correspondence and requests for materials should be addressed to C. Kim (changyoung@snu.ac.kr) and Y. K. Kim (YKim@lbl.gov).

\noindent
\newline
{\bf Competing financial interests}
\newline
The authors declare no competing financial interests.

\vspace{10 pt}

\end{document}